\documentclass[twocolumn,showpacs,preprintnumbers]{revtex4}

\begin{document}

\title{Virtual Compton scattering in the generalized Bjorken region and
positivity bounds on generalized parton distributions}
\author{P.~V.~Pobylitsa}
\affiliation{Institute for Theoretical Physics II, Ruhr University Bochum, D-44780
Bochum, Germany\\
and Petersburg Nuclear Physics Institute, Gatchina, St.~Petersburg, 188300,
Russia}

\begin{abstract}
The positivity bounds on generalized parton distributions are derived from
the positivity properties of the absorptive part of the amplitude of the
virtual Compton scattering in the generalized Bjorken region.
\end{abstract}

\maketitle

\section{Introduction}

Generalized parton distributions (GPDs)
\cite{MRGDH-94,Radyushkin-96-a,Radyushkin-96,Ji-97,Ji-97-b,CFS-97,Radyushkin-97,Radyushkin-review,GPV,BMK-2001,AMS}
appear in the context of the QCD factorization in various hard exclusive
phenomena including deeply virtual Compton scattering and hard exclusive
meson production. Among several general constraints on GPDs an important
role is played by the so-called positivity bounds. Various inequalities for
GPDs have been derived in Refs.~\cite{Martin-98,
Radyushkin-99,PST-99,Ji-98,DFJK-00,Burkardt-01,Pobylitsa-01,Pobylitsa-02,Diehl-02,Burkardt-02-a,Burkardt-02-b}.
As shown in Ref.~\cite{Pobylitsa-02-c}, these inequalities can be
considered as particular cases of a general positivity bound which has a
relatively simple form in the impact parameter representation for GPDs
\cite{Burkardt-01,Diehl-02,Burkardt-02-a,Burkardt-02-b,Burkardt-00}. This
positivity bound is stable under the one-loop evolution of GPDs to higher
normalization points \cite{Pobylitsa-02-c}. The positivity bound of
Ref.~\cite{Pobylitsa-02-c} was explicitly checked for one-loop GPDs in various
perturbative models \cite{Pobylitsa-02-e}. The solutions of this positivity
bound are studied in Refs.~\cite{Pobylitsa-02-e,Pobylitsa-02-d}. The
positivity bounds can be used for self-consistency checks of models of GPDs 
\cite{TM-02}.

The derivation of the positivity bounds on GPDs in Ref.~\cite{Pobylitsa-02-c}
was based on the positivity of the norm in the Hilbert space of states
\begin{equation}
\left\| \sum\limits_{\sigma }\int \frac{dP^{+}d^{2}P^{\perp }d\lambda }{
2P^{+}}g_{\sigma }(\lambda ,P)\psi \left( \lambda n\right) |P,\sigma \rangle
\right\| ^{2}\geq 0\,.  \label{start-ineq}
\end{equation}
Here $|P,\sigma \rangle $ is the hadron state with momentum $P$ and spin
$\sigma $. The quark field $\psi $ is taken at the point $\lambda n$ defined
by the light-cone vector $n$, and the ``good'' spin components of the field
$\psi $ are implied. The superposition of quark-hadron states is weighted
with arbitrary functions $g_{\sigma }$. Expanding the square on the
left-hand side (LHS) of the inequality (\ref{start-ineq}), one obtains the
following inequality:
\[
\sum\limits_{\sigma _{1}\sigma _{2}}\int \frac{dP_{1}^{+}d^{2}P_{1}^{\perp
}d\lambda _{1}}{2P_{1}^{+}}\int \frac{dP_{2}^{+}d^{2}P_{2}^{\perp }d\lambda
_{2}}{2P_{2}^{+}}g_{\sigma _{1}}(\lambda _{1},P_{1}) 
\]
\begin{equation}
\times g_{\sigma _{2}}^{\ast }(\lambda _{2},P_{2})\langle P_{2},\sigma _{2}|
\bar{\psi}\left( \lambda _{2}n\right) (n\gamma )\psi \left( \lambda
_{1}n\right) |P_{1},\sigma _{1}\rangle \geq 0\,.  \label{quark-ineq-2}
\end{equation}
This inequality contains the matrix elements which enter the definition of
GPD. For simplicity we restrict the consideration to the case of spin-0
hadrons. In this case the GPD is defined as follows:
\[
H(x,\xi ,t)=\int \frac{d\lambda }{4\pi }\exp \left\{ \frac{1}{2}ix\lambda 
\left[ n(P_{1}+P_{2})\right] \right\} 
\]
\begin{equation}
\times \langle P_{2}|\bar{\psi}\left( -\frac{\lambda n}{2}\right) (n\gamma
)\psi \left( \frac{\lambda n}{2}\right) |P_{1}\rangle \,.  \label{H-scalar}
\end{equation}
Here we use the standard Ji variables $x,\xi ,t$ \cite{Ji-98}:
\begin{equation}
\xi =\frac{n(P_{1}-P_{2})}{n(P_{1}+P_{2})},\quad t=(P_{2}-P_{1})^{2}\,.
\label{xi-t-Ji}
\end{equation}
Taking into account that inequality (\ref{quark-ineq-2}) should hold for
arbitrary functions $g_{\sigma }$, one can derive positivity bounds on the
GPD $H$ \cite{Pobylitsa-02-c}.

The approach based on inequality (\ref{start-ineq}) raises a number of
questions. The bilinear quark operators have light-cone singularities which
have to be renormalized. The renormalization involves subtractions which can
violate the inequality. Next, the quark-hadron states in the inequality
(\ref{start-ineq}) do not belong to the physical sector and one can wonder
whether the positivity of the norm in the Hilbert space is guaranteed.
Intuitively one can expect that at large normalization points (where the
renormalization is effectively equivalent to the regularization) the
positivity is not destroyed by the renormalization. One can also think that
the insertion of the gluon $P$ exponent between the quark fields in the
inequality (\ref{quark-ineq-2}) will also protect us from the violations of
the positivity of the norm which can be met in the nonphysical sector. On
the other hand, these arguments in favor of positivity are not completely
impeccable and it makes sense to look for an alternative derivation of the
positivity bounds on GPDs which would avoid the problems related to the
light-cone singularities, renormalization and properties of the nonphysical
quark-hadron states.

At this point it is useful to recall the similar problems and their solution
in the context of the forward parton distributions. It is well known that in
the case of the deep inelastic scattering (DIS) one should distinguish
between the \emph{parton distributions} and the \emph{structure functions}.

The \emph{parton distributions} are defined in terms of matrix elements of
quark fields
\begin{equation}
q(x)=\int \frac{d\lambda }{4\pi }e^{i\lambda x(nP)}\langle P|\bar{\psi}
(\lambda n)(n\gamma )\psi \left( 0\right) |P\rangle
\end{equation}
whereas the \emph{structure functions} are extracted from the hadronic tensor
\begin{equation}
W_{\mu \nu }(P,q)=\frac{1}{4\pi }\int d^{4}ze^{iqz}\langle P|j_{\mu
}(z)j_{\nu }(0)|P\rangle \,.
\end{equation}
The positivity of parton distributions is usually associated with their
physical meaning in the infinite momentum frame as the probability to find a
quark with a given momentum fraction. This physical interpretation makes
sense only at large normalization points $\mu $ whereas at low normalization
points the positivity of parton distributions can be violated. In contrast,
the structure functions are directly related to DIS cross sections and the
positivity of cross sections imposes constraints on structure functions,
which hold for any $Q^{2}\equiv -q^{2}$ including the low values of $Q^{2}$.
At large $Q^{2}$ the difference between the structure functions and the
parton distributions (taken at $\mu \sim Q$) can be neglected and in this
case we have common positivity properties for parton distributions and
structure functions.

Now one can ask the question whether this picture of the relations between
the structure functions and parton distributions can be generalized from the
forward case to the case of GPDs? The answer to this question is rather
simple: one should start the analysis from the positivity of the norm the
following state

\begin{equation}
\left\| \int d^{4}z\int \frac{d^{3}P}{2P^{+}}f^{\mu }(z,P)j_{\mu
}(z)|P\rangle \right\| ^{2}\geq 0  \label{f-j-ineq}
\end{equation}
where $j_{\mu }$ is the color singlet quark current
\begin{equation}
j_{\mu }=\bar{\psi}\gamma _{\mu }\psi \,.
\end{equation}
Inequality (\ref{f-j-ineq}) can be rewritten in the form
\[
\int d^{4}z_{1}\int \frac{d^{3}P_{1}}{2P_{1}^{+}}f^{\mu
_{1}}(z_{1},P_{1})\int d^{4}z_{2}\int \frac{d^{3}P_{2}}{2P_{2}^{+}} 
\]
\begin{equation}
\times f^{\mu _{2}\ast }(z_{2},P_{2})\langle P_{2}|j_{\mu _{2}}(z_{2})j_{\mu
_{1}}(z_{1})|P_{1}\rangle \geq 0\,.  \label{f-j-int}
\end{equation}
Here we deal with color singlet currents $j_{\mu }$ and physical hadronic
states. Therefore this inequality should certainly hold for any functions
$f^{\mu }(z,P)$.

Note that the matrix element $\langle P_{2}|j_{\mu _{2}}j_{\mu
_{1}}|P_{1}\rangle $ appearing in the inequality (\ref{f-j-int}) contains
the usual product of currents $j_{\mu }$. In the momentum representation it
can be expressed in terms of the discontinuities of the corresponding matrix
element with the time-ordered product $\langle P_{2}|T\{j_{\mu _{2}}j_{\mu
_{1}}\}|P_{1}\rangle $. These time-ordered matrix elements are directly
related to the ``scattering amplitudes'' involving two virtual photons. In
the so-called generalized Bjorken region \cite{MRGDH-94,BR-00} these
amplitudes can be expressed in terms GPDs.

Similarly, in the case of the hard kinematics, we can express the matrix
elements $\langle P_{2}|j_{\mu _{2}}j_{\mu _{1}}|P_{1}\rangle $ in terms of
the GPDs. Our aim is to choose functions $f^{\mu }(z,P)$ so that the
integral on the LHS of inequality (\ref{f-j-int}) is saturated by the hard
kinematics where the matrix element $\langle P_{2}|j_{\mu _{2}}j_{\mu
_{1}}|P_{1}\rangle $ can be reduced to the GPD. This will lead us to the
positivity bounds on GPDs.

The structure of this paper is as follows. In Section
\ref{positivity-section} we determine kinematical regions relevant for the
derivation of the positivity bounds. In Section \ref{hard-kinematics-section}
we describe the constraints on functions $f^{\mu }$ which allow us to
express matrix elements $\langle P_{2}|j_{\mu _{2}}j_{\mu _{1}}|P_{1}\rangle 
$ in terms of GPDs. The corresponding inequality for GPDs is derived in
Section \ref{reduction-section} where we establish the equivalence of the
approach based on the quark-hadron inequality (\ref{start-ineq}) and of the
current method relying on the positivity property (\ref{f-j-int}) of the
matrix element of currents $j_{\mu }$. The equivalence of the two approaches
is established without using the explicit form of the positivity bound on
GPDs in the impact parameter representation which is briefly described in
Section \ref{impact-representation-section} (the technical details of the
derivation can be found in Appendix \ref{impact-parameter-appendix}).

\section{Positivity bounds for the matrix element of currents}

\label{positivity-section}

In the momentum representation inequality (\ref{f-j-int}) takes the form
\[
\int \frac{d^{3}P_{1}d^{4}q_{1}}{2P_{1}^{+}}h^{\mu _{1}}(P_{1},q_{1})\int 
\frac{d^{3}P_{2}d^{4}q_{2}}{2P_{2}^{+}}h^{\mu _{2}\ast }(P_{2},q_{2}) 
\]
\begin{equation}
\times (2\pi )^{4}\delta ^{(4)}(P_{1}+q_{1}-P_{2}-q_{2})A_{\mu _{2}\mu
_{1}}(q_{1},q_{2};P_{1},P_{2})\geq 0\,.  \label{F-bound}
\end{equation}
Here $h^{\mu }(P,q)$ are arbitrary functions,

\begin{equation}
A_{\mu _{2}\mu _{1}}(q_{1},q_{2};P_{1},P_{2})=\int d^{4}ze^{iq_{2}z}\langle
P_{2}|j_{\mu _{2}}\left( z\right) j_{\mu _{1}}\left( 0\right) |P_{1}\rangle
\end{equation}
and the momentum conservation reads:
\begin{equation}
P_{1}+q_{1}=P_{2}+q_{2}\,.  \label{momentum-conservation}
\end{equation}

Let us separate the connected part of the matrix element
\[
\langle P_{2}|j_{\mu _{2}}\left( z_{2}\right) j_{\mu _{1}}\left(
z_{1}\right) |P_{1}\rangle ^{conn} 
\]
\begin{equation}
\equiv \langle P_{2}|j_{\mu _{2}}\left( z_{2}\right) j_{\mu _{1}}\left(
z_{1}\right) |P_{1}\rangle -\langle P_{2}|P_{1}\rangle \langle 0|j_{\mu
_{2}}\left( z_{2}\right) j_{\mu _{1}}\left( z_{1}\right) |0\rangle
\end{equation}
and define
\[
A_{\mu _{2}\mu _{1}}^{conn}(q_{1},q_{2};P_{1},P_{2}) 
\]
\begin{equation}
=\int d^{4}ze^{iq_{2}z}\langle P_{2}|j_{\mu _{2}}\left( z\right) j_{\mu
_{1}}\left( 0\right) |P_{1}\rangle ^{conn}\,.  \label{A-connected}
\end{equation}
Note that in the hard limit only the connected part $A_{\mu _{2}\mu
_{1}}^{conn}$ reduces to the GPD. In order to avoid the contamination of
inequality (\ref{F-bound}) by the vacuum part $\langle 0|j_{\mu _{2}}j_{\mu
_{1}}|0\rangle $ we must choose functions $h^{\mu }$ so that the vacuum part
does not contribute. Let $m_{0}$ be the mass of the lightest intermediate
state contributing to $\langle 0|j_{\mu _{2}}j_{\mu _{1}}|0\rangle $. The
vacuum part $\langle 0|j_{\mu _{2}}j_{\mu _{1}}|0\rangle $ will be
suppressed if for 
\begin{equation}
q^{2}\geq m_{0}^{2},\quad q^{0}>0\,  \label{allowed-q}
\end{equation}
we have
\begin{equation}
h^{\mu }(P,q)=0\,.
\end{equation}

Next, the matrix element $A_{\mu _{2}\mu _{1}}(q_{1},q_{2};P_{1},P_{2})$
does not vanish only in a certain kinematical region of variables
$P_{i},q_{i}$. Let us assume for simplicity that the lightest intermediate
state $|n\rangle $, which can contribute to the matrix element
\[
\langle P_{2}|j_{\mu _{2}}\left( z_{2}\right) j_{\mu _{1}}\left(
z_{1}\right) |P_{1}\rangle 
\]
\begin{equation}
=\sum\limits_{n}\langle P_{2}|j_{\mu _{2}}\left( z_{2}\right) |n\rangle
\langle n|j_{\mu _{1}}\left( z_{1}\right) |P_{1}\rangle \,,
\end{equation}
has the same mass as $P_{1}$ and $P_{2}$. Then the following condition is
necessary if one wants to have nonvanishing $A_{\mu _{2}\mu
_{1}}(q_{1},q_{2};P_{1},P_{2})$:
\begin{equation}
2P_{1}q_{1}+q_{1}^{2}\geq 0\,,\quad (P_{1}+q_{1})^{0}\geq 0\,.
\label{nonzero-condition}
\end{equation}
Excluding the region (\ref{allowed-q}), we see that we must deal with
functions $h^{\mu }(P,q)$ vanishing if at least one of the following
conditions holds:
\begin{equation}
h^{\mu }(P,q)=0\quad \left\{ 
\begin{array}{l}
\mathrm{if}\quad q^{2}\geq m_{0}^{2}\quad \mathrm{and}\quad q^{0}\geq 0 \\ 
\mathrm{if}\quad 2Pq+q^{2}\leq 0 \\ 
\mathrm{if}\quad (P+q)^{0}\leq 0
\end{array}
\right. \,
\end{equation}
Obviously we can replace the zero components by the projections on any time-
or light-like vector $n$ (with $n^{0}>0$):
\begin{equation}
h^{\mu }(P,q)=0\quad \left\{ 
\begin{array}{l}
\mathrm{if}\quad q^{2}\geq m_{0}^{2}\quad \mathrm{and}\quad nq\geq 0 \\ 
\mathrm{if}\quad 2Pq+q^{2}\leq 0 \\ 
\mathrm{if}\quad n(P+q)\leq 0
\end{array}
\right. \,  \label{f-constraint}
\end{equation}

\section{Hard kinematics}

\label{hard-kinematics-section}

We want to choose functions $h^{\mu }(P,q)$ so that matrix elements $A_{\mu
_{2}\mu _{1}}(q_{1},q_{2};P_{1},P_{2})$ appear in the inequality
(\ref{F-bound}) only in the hard kinematics (corresponding to the generalized
Bjorken region of Refs. \cite{MRGDH-94,BR-00})
\begin{equation}
-q_{k}^{2}\sim (P_{i}q_{n})\gg (P_{i}P_{n})\sim \Lambda _{QCD}^{2}\,.
\label{hard-limit}
\end{equation}
This kinematics will allow us to reduce the matrix element $A_{\mu _{2}\mu
_{1}}$ to the GPDs and to derive the positivity bounds on GPDs from the
inequality (\ref{F-bound}). To this aim we take
\begin{equation}
q_{i}=\rho n-k_{i}\quad (i=1,2)  \label{q-hard}
\end{equation}
where $n$ is a light-light-cone vector
\begin{equation}
n^{2}=0
\end{equation}
with the positive time component $n^{0}>0$ so that
\begin{equation}
(nP_{1}),\quad (nP_{2})>0.  \label{nP-positive}
\end{equation}
Parameter $\rho $ is assumed to be large: 
\begin{equation}
\rho \rightarrow \infty  \label{rho-large}
\end{equation}
and momenta
\begin{equation}
k_{i},P_{i}=\mathrm{const}
\end{equation}
are fixed in this limit. In this hard limit the constraint
(\ref{f-constraint}) takes the form
\begin{equation}
h^{\mu }(P,\rho n-k)=0\quad \left\{ 
\begin{array}{l}
\mathrm{if}\quad \rho (nk)\leq 0\quad \mathrm{and}\quad (nk)\leq 0 \\ 
\mathrm{if}\quad \rho n(P-k)\leq 0 \\ 
\mathrm{if}\quad n(P-k)\leq 0
\end{array}
\right.
\end{equation}
If $\rho <0$ then the last two lines lead to the completely vanishing
function $h^{\mu }=0$. Therefore the limit $\rho \rightarrow \infty $ should
be understood as $\rho \rightarrow +\infty $. In this case the above
constraints on $h^{\mu }$ simplify as follows:
\begin{equation}
h^{\mu }(P,\rho n-k)=0\quad \left\{ 
\begin{array}{l}
\mathrm{if}\quad (nk)\leq 0 \\ 
\mathrm{if}\quad n(P-k)\leq 0
\end{array}
\right.
\end{equation}
Taking into account condition (\ref{nP-positive}) we conclude that
\begin{equation}
h^{\mu }(P,\rho n-k)\neq 0\quad \mathrm{only\,if}\quad 0\leq (nk)\leq (nP)\,.
\label{f-constraint-2}
\end{equation}
Certainly we also assume that $h^{\mu }(P,\rho n-k)=0$ if $\rho $ is not
large enough.

Let us introduce the following kinematical variables:
\begin{equation}
x=-\frac{q_{2}^{2}+q_{1}^{2}}{(P_{2}+P_{1})(q_{1}+q_{2})}\,,
\end{equation}
\begin{equation}
\xi =\frac{(P_{1}-P_{2})(q_{1}+q_{2})}{(P_{2}+P_{1})(q_{1}+q_{2})}\,.
\end{equation}
The choice of notation for these variables is motivated by the compatibility
with Ji variables $x,\xi $ [see Eqs. (\ref{H-scalar}), (\ref{xi-t-Ji})] in
the hard limit.

In the hard limit (\ref{q-hard}), (\ref{rho-large}) we have
\begin{equation}
x=\frac{n(k_{1}+k_{2})}{n(P_{1}+P_{2})}\,,  \label{x-hard}
\end{equation}
\begin{equation}
\xi =\frac{n(P_{1}-P_{2})}{n(P_{1}+P_{2})}=\frac{n(k_{1}-k_{2})}{
n(P_{1}+P_{2})}.  \label{xi-hard}
\end{equation}
The property (\ref{f-constraint-2}) of functions $h^{\mu }$ guarantees that
in the integral (\ref{F-bound}) we deal only with the case
\begin{equation}
0\leq (nk_{i})\leq (nP_{i})\,.
\end{equation}
This means that the corresponding parameters $x,\xi $ are constrained to the
following region:
\begin{equation}
|\xi |\leq x\leq 1\,.  \label{xi-x-region}
\end{equation}

\section{Reduction of the matrix element of currents to GPD}

\label{reduction-section}

The next step is to notice that the matrix element $A_{\mu _{2}\mu
_{1}}^{conn}$ (\ref{A-connected}) can be reduced to GPD (\ref{H-scalar}) in
the hard limit (\ref{q-hard}), (\ref{rho-large}). Indeed, in this limit we
have 
\[
A_{\mu _{2}\mu _{1}}^{conn}(q_{1},q_{2};P_{1},P_{2}) 
\]
\[
=\int d^{4}ze^{iq_{2}z}\langle P_{2}|j_{\mu _{2}}\left( z\right) j_{\mu
_{1}}\left( 0\right) |P_{1}\rangle ^{conn} 
\]
\[
\rightarrow \frac{1}{2}\int d\lambda \exp \left\{ \frac{1}{2}ix\lambda \left[
n(P_{1}+P_{2})\right] \right\} 
\]
\[
\times \langle P_{2}|\bar{\psi}\left( -\frac{\lambda n}{2}\right) \gamma
_{\mu _{2}}(n\gamma )\gamma _{\mu _{1}}\psi \left( \frac{\lambda n}{2}
\right) |P_{1}\rangle 
\]
\[
-\frac{1}{2}\int d\lambda \exp \left\{ -\frac{1}{2}ix\lambda \left[
n(P_{1}+P_{2})\right] \right\} 
\]
\begin{equation}
\times \langle P_{2}|\bar{\psi}\left( -\frac{\lambda n}{2}\right) \gamma
_{\mu _{1}}(n\gamma )\gamma _{\mu _{2}}\psi \left( \frac{\lambda n}{2}
\right) |P_{1}\rangle \,.  \label{P-jj-P-factorization}
\end{equation}
This expression can be obtained by calculating the discontinuities of the
time-ordered matrix elements studied in Ref.~\cite{MRGDH-94}. Alternatively
one can derive Eq.~(\ref{P-jj-P-factorization}) by treating the quark fields
as free and using Wick theorem.

Let us introduce a light-cone vector $p$ dual to $n$
\begin{equation}
p^{2}=0\,,\quad (pn)\neq 0\,.
\end{equation}
Using the light-cone vectors $p$ and $n$, we define the projector
\begin{equation}
\Pi _{\mu \nu }=g_{\mu \nu }-\frac{1}{(pn)}\left( p_{\mu }n_{\nu }+n_{\mu
}p_{\nu }\right)
\end{equation}
with the properties
\begin{equation}
\Pi _{\mu \nu }\Pi ^{\nu \rho }=\delta _{\mu }^{\rho }\,,
\end{equation}
\begin{equation}
\Pi _{\mu \nu }n^{\nu }=0\,,\quad \Pi _{\mu \nu }p^{\nu }=0\,\,.
\end{equation}
Obviously $\Pi _{\mu \nu }$ is a projector on the transverse plane where one
can choose basis $e^{(1)},$ $e^{(2)}$:
\begin{equation}
-\Pi _{\mu \nu }=\sum\limits_{a=1,2}e_{\mu }^{(a)}e_{\nu }^{(a)},
\label{Pi-ee}
\end{equation}
\begin{equation}
e^{(1)}e^{(1)}=e^{(2)}e^{(2)}=-1,
\end{equation}
\begin{equation}
(e^{a}n)=(e^{a}p)=0\,.
\end{equation}
Taking into account that
\begin{equation}
-\Pi ^{\mu _{2}\mu _{1}}\gamma _{\mu _{2}}(n\gamma )\gamma _{\mu
_{1}}=2(n\gamma )\,,
\end{equation}
we find from Eq.~(\ref{P-jj-P-factorization}) in the hard limit
\[
\frac{1}{4\pi }\sum\limits_{a=1,2}\int d^{4}ze^{iq_{2}z}\langle P_{2}|\left[
e^{a}\cdot j\left( z\right) \right] \left[ e^{a}\cdot j\left( 0\right) 
\right] |P_{1}\rangle 
\]
\begin{equation}
\rightarrow H(x,\xi ,t)-H(-x,\xi ,t)\,.  \label{jj-GPD-equiv}
\end{equation}
The GPD $H(x,\xi ,t)$ was defined in Eq.~(\ref{H-scalar}). The LHS of
Eq.~(\ref{jj-GPD-equiv}) obeys the positivity bound (\ref{F-bound}). Taking in
inequality (\ref{F-bound})
\begin{equation}
h^{\mu }(P,\rho n-k)\equiv e^{(a)\mu }s(P,k)\,
\end{equation}
and summing over $a=1,2$, we derive using Eq.~(\ref{jj-GPD-equiv})
\[
\int \frac{d^{3}P_{1}d^{4}k_{1}}{2P_{1}^{+}}s(P_{1},k_{1})\int \frac{
d^{3}P_{2}d^{4}k_{2}}{2P_{2}^{+}}s^{\ast }(P_{2},k_{2}) 
\]
\[
\times (2\pi )^{4}\delta ^{(4)}(P_{1}-k_{1}-P_{2}+k_{2}) 
\]
\begin{equation}
\times \left[ H(x,\xi ,t)-H(-x,\xi ,t)\right] \geq 0\,.  \label{H-ineq-1}
\end{equation}
The variables $x,\xi $ on the right-hand side are assumed to be expressed in
terms of $P_{i},k_{i}$ according to Eqs. (\ref{x-hard}), (\ref{xi-hard}).
Function $s(P,k)$ is arbitrary up to the constraint (\ref{f-constraint-2}):
\begin{equation}
s(P,k)\neq 0\quad \mathrm{only\,if}\quad 0\leq (nk)\leq (nP)\,.
\label{s-region}
\end{equation}
This means that inequality (\ref{H-ineq-1}) covers only the region $|\xi
|\leq x\leq 1$ (\ref{xi-x-region}). The GPD $H(x,\xi ,t)$ should be taken at
the normalization point $\mu $ determined (with the leading logarithm
accuracy) by the hard scale (\ref{hard-limit})

\begin{equation}
\mu ^{2}\sim -q_{k}^{2}\gg -t\sim \Lambda _{QCD}^{2}\,.  \label{limits}
\end{equation}

Note that the inequality (\ref{H-ineq-1}) contains two terms
\begin{equation}
H(x,\xi ,t)-H(-x,\xi ,t)\,.
\end{equation}
At $x>|\xi |$ the first term $H(x,\xi ,t)$ can be interpreted as the quark
contribution whereas $-H(-x,\xi ,t)$ corresponds to the antiquarks. Actually
inequality (\ref{H-ineq-1}) is a sum of two independent positivity bounds
for the quark and antiquark distributions 
\[
\int \frac{d^{3}P_{1}d^{4}k_{1}}{2P_{1}^{+}}s(P_{1},k_{1})\int \frac{
d^{3}P_{2}d^{4}k_{2}}{2P_{2}^{+}}s^{\ast }(P_{2},k_{2}) 
\]
\begin{equation}
\times (2\pi )^{4}\delta ^{(4)}(P_{1}-k_{1}-P_{2}+k_{2})\left[ \pm H(\pm
x,\xi ,t)\right] \geq 0\,.  \label{H-ineq-q}
\end{equation}
The reason, why the positivity bounds for quarks and antiquarks mix in
inequality (\ref{H-ineq-1}), can be understood already at the level of the
forward parton distributions: it is well known that in the electromagnetic
DIS the structure functions contain the sum of the quark and antiquark
distributions with squared electric charges so that for any flavor the quark
and antiquark distributions appear together with the same weight.

In order to separate the quark contribution from the antiquark part, one can
consider the positivity properties of the \emph{left} currents. This is done
in Appendix \ref{Left-currents-appendix} where inequality (\ref{H-ineq-q})
is derived.

Inequality (\ref{H-ineq-q}) can be easily reduced to inequality
(\ref{quark-ineq-2}) which was used as a starting point for the
derivation of the positivity bounds on GPDs in Ref.~\cite{Pobylitsa-02-c}.
Thus we see that the current approach (based on the positivity properties of
matrix elements of currents) and the method of Ref.~\cite{Pobylitsa-02-c}
(relying on the positivity of the norm of the quark-hadron states) lead to
the same bounds on GPDs.

\section{Positivity bounds in the impact parameter representation}

\label{impact-representation-section}

In the previous section it was explained that the quark-hadron method of
Ref.~\cite{Pobylitsa-02-c} and the current approach lead to the same result.
In Ref.~\cite{Pobylitsa-02-c} it was shown that the positivity bounds can be
simplified by using the impact parameter representation for GPDs. In this
section we present only the result. The technical details can be found in
Appendix \ref{impact-parameter-appendix}.

In the frame, where $(P_{1}+P_{2})^{\perp }=0$ and $n^{\perp }=0$, the
transverse component of the transferred momentum $\Delta ^{\perp
}=P_{2}^{\perp }-P_{1}^{\perp }$ is connected with the variable $t$
(\ref{xi-t-Ji}) by the following relation:
\begin{equation}
t=-\frac{|\Delta ^{\perp }|^{2}+4\xi ^{2}M^{2}}{1-\xi ^{2}}\,.
\end{equation}

Let us define the GPD in the impact parameter representation via the Fourier
transformation in $\Delta ^{\perp }$:
\[
\tilde{F}\left( x,\xi ,b^{\perp }\right) =\int \frac{d^{2}\Delta ^{\perp }}{
(2\pi )^{2}}\exp \left[ i(\Delta ^{\perp }b^{\perp })\right] 
\]
\begin{equation}
\times H\left( x,\xi ,-\frac{|\Delta ^{\perp }|^{2}+4\xi ^{2}M^{2}}{1-\xi
^{2}}\right) \,.  \label{F-tilde-def}
\end{equation}
In Appendix \ref{impact-parameter-appendix} the following inequality is
derived from inequality (\ref{H-ineq-q}): 
\[
\,\int\limits_{-1}^{1}d\xi \int\limits_{|\xi |}^{1}\frac{dx}{(1-x)^{5}}
p^{\ast }\left( \frac{1-x}{1-\xi }\right) p\left( \frac{1-x}{1+\xi }\right) 
\]
\begin{equation}
\times \left[ \pm \tilde{F}\left( \pm x,\xi ,\frac{1-x}{1-\xi ^{2}}b^{\perp
}\right) \right] \geq 0\,.  \label{ineq-res}
\end{equation}
This inequality should hold for any function $p$. It coincides with the
positivity bound derived in Ref.~\cite{Pobylitsa-02-c}.

\section{Conclusions}

In this paper the alternative derivation of the positivity bounds on GPDs is
considered. The advantage of this method is that it is based on quite
transparent positivity properties of matrix elements of color singlet
currents over physical hadronic states. This allows us to avoid a number of
problems which arise in the original derivation of the positivity bounds
based on the positivity properties of the nonphysical quark-hadronic states.
From this point of view the derivation of the positivity bounds described in
this paper is more favorable. Another advantage of the new derivation is
that it makes clear certain physical restrictions on the region where the
positivity bounds should hold. We see from Eq.~(\ref{limits}) that the
normalization point $\mu $ should be large not only compared to $\Lambda
_{QCD}$ but also the condition $\mu ^{2}\gg |t|$ should hold. In terms of
the impact parameter $b^{\perp }$ used in the explicit formulation of the
positivity bounds (\ref{ineq-res}) this means that for the validity of the
positivity bounds we need the condition $\mu \gg |b^{\perp }|^{-1}$.

For simplicity our analysis was restricted to the case of spin-0 hadrons.
The generalization to hadrons with nonzero spins is straightforward and the
explicit form of the corresponding positivity bounds can be found in
Ref.~\cite{Pobylitsa-02-c}.

\textbf{Acknowledgement.} I am grateful to Ya.~I.~Azimov and M.~V.~Polyakov
for useful discussions.

\appendix

\section{Positivity bounds on GPDs and left currents}

\label{Left-currents-appendix}

In this appendix we derive inequality (\ref{H-ineq-q}) using the positivity
properties of the matrix element of the left currents
\begin{equation}
j_{\mu }^{L}=\bar{\psi}\gamma _{\mu }(1-\gamma _{5})\psi \,.
\end{equation}
By analogy with Eq.~(\ref{P-jj-P-factorization}) we have in the hard limit
\[
\int d^{4}ze^{iq_{2}z}\langle P_{2}|j_{\mu _{2}}^{L\dagger }(z)j_{\mu
_{1}}^{L}(0)|P_{1}\rangle 
\]
\[
\rightarrow \int d\lambda \exp \left\{ \frac{1}{2}ix\lambda \left[
n(P_{1}+P_{2})\right] \right\} 
\]
\[
\times \langle P_{2}|\bar{\psi}\left( -\frac{\lambda n}{2}\right) \gamma
_{\mu _{2}}(n\gamma )\gamma _{\mu _{1}}(1-\gamma _{5})\psi \left( \frac{
\lambda n}{2}\right) |P_{1}\rangle 
\]
\[
-\int d\lambda \exp \left\{ -\frac{1}{2}ix\lambda \left[ n(P_{1}+P_{2})
\right] \right\} 
\]
\begin{equation}
\times \langle P_{2}|\bar{\psi}\left( -\frac{\lambda n}{2}\right) \gamma
_{\mu _{1}}(n\gamma )\gamma _{\mu _{2}}(1-\gamma _{5})\psi \left( \frac{
\lambda n}{2}\right) |P_{1}\rangle \,.  \label{ME-left-calc}
\end{equation}
Let us introduce the vector describing the helicity eigenstate of the
virtual photon 
\begin{equation}
e=\frac{1}{\sqrt{2}}\left( e^{(1)}+ie^{(2)}\right) 
\end{equation}
with the properties
\begin{equation}
(en)=0,\quad (ee)=0\,,\quad (e^{\ast }e)=-1\,,
\end{equation}
\begin{equation}
\varepsilon _{\lambda \mu \nu \rho }e^{\lambda \ast }e^{\mu }n^{\nu
}=-in_{\rho }\,.
\end{equation}
Using relations
\begin{equation}
(e^{\ast }\gamma )(e\gamma )(n\gamma )=-(1+\gamma _{5})(n\gamma )\,,
\label{eps-def}
\end{equation}
\begin{equation}
(e\gamma )(e^{\ast }\gamma )(n\gamma )=-(1-\gamma _{5})(n\gamma ),
\end{equation}
we find from Eq.~(\ref{ME-left-calc}) that in the hard limit
\[
\int d^{4}ze^{iq_{2}z}\langle P_{2}|\left[ e^{\mu _{2}}j_{\mu _{2}}^{L}(z)
\right] ^{+}\left[ e^{\mu _{1}}j_{\mu _{1}}^{L}(0)\right] |P_{1}\rangle 
\]
\[
\rightarrow 2\int d\lambda \exp \left\{ \frac{1}{2}ix\lambda \left[
n(P_{1}+P_{2})\right] \right\} 
\]
\[
\times \langle P_{2}|\bar{\psi}\left( -\frac{\lambda n}{2}\right) (n\gamma
)\psi \left( \frac{\lambda n}{2}\right) |P_{1}\rangle 
\]
\begin{equation}
=8\pi H(x,\xi ,t)\,.  \label{jj-L-hard}
\end{equation}
Using this relation instead of Eq.~(\ref{jj-GPD-equiv}), we obtain
inequality (\ref{H-ineq-q}) with the upper sign choice in $\pm $ by analogy
with the derivation of inequality (\ref{H-ineq-1}).

Replacing $e\rightarrow e^{\ast }$ in the LHS of Eq.~(\ref{jj-L-hard}) we
find
\[
\int d^{4}ze^{iq_{2}z}\langle P_{2}|\left[ e^{\mu _{2}\ast }j_{\mu
_{2}}^{L}(z)\right] ^{+}\left[ e^{\mu _{1}\ast }j_{\mu _{1}}^{L}(0)\right]
|P_{1}\rangle 
\]
\[
\rightarrow -2\int d\lambda \exp \left\{ -\frac{1}{2}ix\lambda \left[
n(P_{1}+P_{2})\right] \right\} 
\]
\[
\times \langle P_{2}|\bar{\psi}\left( -\frac{\lambda n}{2}\right) (n\gamma
)\psi \left( \frac{\lambda n}{2}\right) |P_{1}\rangle 
\]
\begin{equation}
=-8\pi H(-x,\xi ,t)\,.
\end{equation}
This result allows us to derive inequality (\ref{H-ineq-q}) with the minus
sign choice.

\section{Derivation of positivity bounds on GPDs in the impact parameter
representation}

\label{impact-parameter-appendix}

In this appendix we derive the positivity bound in the impact parameter
representation (\ref{ineq-res}) from the inequality (\ref{H-ineq-q}).

Let us choose the light-cone coordinates so that for any vector $X^{\mu }$
\begin{equation}
X^{+}=\mathrm{const\,}(nX)\,.
\end{equation}
Then according to Eqs. (\ref{x-hard}), (\ref{xi-hard}) 
\begin{equation}
x=\frac{k_{1}^{+}+k_{2}^{+}}{P_{1}^{+}+P_{2}^{+}},\quad \xi =\frac{
P_{1}^{+}-P_{2}^{+}}{P_{1}^{+}+P_{2}^{+}}\,.  \label{xi-x-LC}
\end{equation}
We can rewrite inequality (\ref{H-ineq-1}) in the following form (in the
case of the upper sign $\pm $)
\[
\int \frac{d^{3}P_{1}d^{4}k_{1}}{2P_{1}^{+}}s(P_{1},k_{1})\int \frac{
d^{3}P_{2}d^{4}k_{2}}{2P_{2}^{+}}s^{\ast }(P_{2},k_{2})\,. 
\]
\[
\times (2\pi )^{4}\delta ^{(4)}(P_{1}-k_{1}-P_{2}+k_{2}) 
\]
\begin{equation}
\times H\left[ \frac{k_{1}^{+}+k_{2}^{+}}{P_{1}^{+}+P_{2}^{+}},\frac{
P_{1}^{+}-P_{2}^{+}}{P_{1}^{+}+P_{2}^{+}},(P_{2}-P_{1})^{2}\right] \geq 0\,.
\label{H-ineq-2}
\end{equation}
Using the Fourier representation for the delta function
\[
2\pi \delta (P_{1}^{-}-k_{1}^{-}-P_{2}^{-}+k_{2}^{-})=\int dy 
\]
\begin{equation}
\times \exp \left[ iy\left( \frac{|P_{1}^{\perp }|^{2}+M^{2}}{2P_{1}^{+}}
-k_{1}^{-}-\frac{|P_{2}^{\perp }|^{2}+M^{2}}{2P_{2}^{+}}+k_{2}^{-}\right) 
\right] \,,
\end{equation}
we can reduce inequality (\ref{H-ineq-2}) to the form
\[
\int dy\int \frac{d^{3}P_{1}d^{3}k_{1}}{2P_{1}^{+}}
s_{2}(P_{1},k_{1}^{+},k_{1}^{\perp },y)\int \frac{d^{3}P_{2}d^{3}k_{2}}{
2P_{2}^{+}} 
\]
\[
\times s_{2}^{\ast }(P_{2},k_{2}^{+},k_{2}^{\perp },y)(2\pi )^{3}\delta
(P_{1}^{+}-k_{1}^{+}-P_{2}^{+}+k_{2}^{+}) 
\]
\[
\times \delta ^{(2)}(P_{1}^{\perp }-k_{1}^{\perp }-P_{2}^{\perp
}+k_{2}^{\perp }) 
\]
\begin{equation}
\times H\left[ \frac{k_{1}^{+}+k_{2}^{+}}{P_{1}^{+}+P_{2}^{+}},\frac{
P_{1}^{+}-P_{2}^{+}}{P_{1}^{+}+P_{2}^{+}},(P_{2}-P_{1})^{2}\right] \geq 0\,,
\label{H-ineq-3}
\end{equation}
where
\[
s_{2}(P,k^{+},k^{\perp },y)=\int \frac{dk^{-}}{2\pi }s(P,k) 
\]
\begin{equation}
\times \exp \left[ iy\left( \frac{|P^{\perp }|^{2}+M^{2}}{2P^{+}}
-k^{-}\right) \right] \,.
\end{equation}
In inequality (\ref{H-ineq-3}) the dependence of functions
$s_{2}(P,k^{+},k^{\perp },y)$ on $y$ is arbitrary. Therefore inequality
(\ref{H-ineq-3}) should hold before the integration over $y$ for any fixed $y$.
Thus we conclude that for any function $s_{3}(P,k^{+},k^{\perp })$
\[
\int \frac{d^{3}P_{1}d^{3}k_{1}}{2P_{1}^{+}}s_{3}(P_{1},k_{1}^{+},k_{1}^{
\perp })\int \frac{d^{3}P_{2}d^{3}k_{2}}{2P_{2}^{+}}s_{3}^{\ast
}(P_{2},k_{2}^{+},k_{2}^{\perp }) 
\]
\[
\times (2\pi )^{3}\delta (P_{1}^{+}-k_{1}^{+}-P_{2}^{+}+k_{2}^{+})\delta
^{(2)}(P_{1}^{\perp }-k_{1}^{\perp }-P_{2}^{\perp }+k_{2}^{\perp }) 
\]
\begin{equation}
\times H\left[ \frac{k_{1}^{+}+k_{2}^{+}}{P_{1}^{+}+P_{2}^{+}},\frac{
P_{1}^{+}-P_{2}^{+}}{P_{1}^{+}+P_{2}^{+}},(P_{2}-P_{1})^{2}\right] \geq 0\,.
\label{H-ineq-4}
\end{equation}
Similarly we define the function
\[
s_{4}(P,k^{+},y^{\perp })=\int \frac{d^{2}k^{\perp }}{(2\pi )^{2}}
s_{3}(P,k^{+},k^{\perp }) 
\]
\begin{equation}
\times \exp \left[ iy^{\perp }(P^{\perp }-k^{\perp })\right] \,.
\end{equation}
Then we find from the inequality (\ref{H-ineq-4})
\[
\int d^{2}y^{\perp }\int \frac{d^{3}P_{1}dk_{1}^{+}}{2P_{1}^{+}}
s_{4}(P_{1},k_{1}^{+},y^{\perp })\int \frac{d^{3}P_{2}dk_{2}^{+}}{2P_{2}^{+}}
\]
\[
\times s_{4}^{\ast }(P_{2},k_{2}^{+},y^{\perp })2\pi \delta
(P_{1}^{+}-k_{1}^{+}-P_{2}^{+}+k_{2}^{+}) 
\]
\begin{equation}
\times H\left[ \frac{k_{1}^{+}+k_{2}^{+}}{P_{1}^{+}+P_{2}^{+}},\frac{
P_{1}^{+}-P_{2}^{+}}{P_{1}^{+}+P_{2}^{+}},(P_{2}-P_{1})^{2}\right] \geq 0\,.
\label{H-ineq-5}
\end{equation}
Again, due to the arbitrary dependence of the function
$s_{4}(P_{1},k_{1}^{+},y^{\perp })$ on $y^{\perp }$, this inequality should
hold before the integration over $y^{\perp }$ for any value of $y^{\perp }$,
i.e. for any function $s_{5}(P,k^{+})$ we must have
\[
\int \frac{d^{3}P_{1}dk_{1}^{+}}{2P_{1}^{+}}s_{5}(P_{1},k_{1}^{+})\int \frac{
d^{3}P_{2}dk_{2}^{+}}{2P_{2}^{+}}s_{5}^{\ast }(P_{2},k_{2}^{+}) 
\]
\[
\times 2\pi \delta (P_{1}^{+}-k_{1}^{+}-P_{2}^{+}+k_{2}^{+}) 
\]
\begin{equation}
\times H\left[ \frac{k_{1}^{+}+k_{2}^{+}}{P_{1}^{+}+P_{2}^{+}},\frac{
P_{1}^{+}-P_{2}^{+}}{P_{1}^{+}+P_{2}^{+}},(P_{2}-P_{1})^{2}\right] \geq 0\,.
\label{H-ineq-6}
\end{equation}
The next step is to notice that
\[
|P_{1}^{+}P_{2}^{\perp }-P_{2}^{+}P_{1}^{\perp }|^{2} 
\]
\begin{equation}
=-P_{1}^{+}P_{2}^{+}(P_{1}-P_{2})^{2}-M^{2}(P_{1}^{+}-P_{2}^{+})^{2}\,.
\label{P-sqr-identity}
\end{equation}
Taking into account the expression (\ref{xi-x-LC}) for $\xi $, we find from
Eq.~(\ref{P-sqr-identity})
\begin{equation}
\left| \frac{2(P_{1}^{+}P_{2}^{\perp }-P_{2}^{+}P_{1}^{\perp })}{
P_{1}^{+}+P_{2}^{+}}\right| ^{2}=(1-\xi ^{2})t-4\xi ^{2}M^{2}\,.
\end{equation}
Let us introduce notation
\begin{equation}
\tilde{\Delta}^{\perp }=\frac{2(P_{1}^{+}P_{2}^{\perp
}-P_{2}^{+}P_{1}^{\perp })}{P_{1}^{+}+P_{2}^{+}}\,.  \label{Delta-tilde}
\end{equation}
Then the variable $t$ (\ref{xi-t-Ji}) is equal to

\begin{equation}
t=-\frac{|\tilde{\Delta}^{\perp }|^{2}+4\xi ^{2}M^{2}}{1-\xi ^{2}}\,.
\end{equation}
Note that in the frame where $(P_{1}+P_{2})^{\perp }=0$, we have
$\tilde{\Delta}^{\perp }=(P_{2}-P_{1})^{\perp }\equiv \Delta ^{\perp }$. However,
here we deal with inequalities containing the integration over $P_{1},P_{2}$
so that the constraint $(P_{1}+P_{2})^{\perp }=0$ cannot be imposed. In
order to avoid confusion with $\Delta ^{\perp }=(P_{2}-P_{1})^{\perp }$, we
use notation $\tilde{\Delta}^{\perp }$ for the variable (\ref{Delta-tilde}).

Now we introduce the function 
\begin{equation}
F(x,\xi ,\tilde{\Delta}^{\perp })=H\left( x,\xi ,\frac{|\tilde{\Delta}
^{\perp }|^{2}+4\xi ^{2}M^{2}}{1-\xi ^{2}}\right) \,.
\end{equation}
In terms of this function inequality (\ref{H-ineq-6}) takes the form
\[
\int \frac{d^{3}P_{1}dk_{1}^{+}}{2P_{1}^{+}}s_{5}(P_{1},k_{1}^{+})\int \frac{
d^{3}P_{2}dk_{2}^{+}}{2P_{2}^{+}}s_{5}^{\ast }(P_{2},k_{2}^{+}) 
\]
\[
\times 2\pi \delta (P_{1}^{+}-k_{1}^{+}-P_{2}^{+}+k_{2}^{+}) 
\]
\begin{equation}
\times F\left[ \frac{k_{1}^{+}+k_{2}^{+}}{P_{1}^{+}+P_{2}^{+}},\frac{
P_{1}^{+}-P_{2}^{+}}{P_{1}^{+}+P_{2}^{+}},\frac{2(P_{1}^{+}P_{2}^{\perp
}-P_{2}^{+}P_{1}^{\perp })}{P_{1}^{+}+P_{2}^{+}}\right] \geq 0\,.
\label{ineq-1}
\end{equation}
Taking functions $s_{5}$ in the factorized form 
\begin{equation}
s_{5}(k^{+},P^{+},P^{\perp })=s_{6}(k^{+},P^{+})s_{7}\left( \frac{P^{\perp }
}{P^{+}}\right) \,,
\end{equation}
we find
\[
\int d^{2}P_{1}^{\perp }\int d^{2}P_{2}^{\perp }s_{7}\left( \frac{
P_{1}^{\perp }}{P_{1}^{+}}\right) s_{7}^{\ast }\left( \frac{P_{2}^{\perp }}{
P_{2}^{+}}\right) 
\]
\[
\times F\left[ \frac{k_{1}^{+}+k_{2}^{+}}{P_{1}^{+}+P_{2}^{+}},\frac{
P_{1}^{+}-P_{2}^{+}}{P_{1}^{+}+P_{2}^{+}},\frac{2(P_{1}^{+}P_{2}^{\perp
}-P_{2}^{+}P_{1}^{\perp })}{P_{1}^{+}+P_{2}^{+}}\right] 
\]
\[
=\frac{1}{4}(P_{1}^{+}+P_{2}^{+})^{2}\int d^{2}b^{\perp }\left| \tilde{s}
_{7}(b^{\perp })\right| ^{2} 
\]
\begin{equation}
\times \tilde{F}\left[ \frac{k_{1}^{+}+k_{2}^{+}}{P_{1}^{+}+P_{2}^{+}},\frac{
P_{1}^{+}-P_{2}^{+}}{P_{1}^{+}+P_{2}^{+}},\frac{P_{1}^{+}+P_{2}^{+}}{
2P_{1}^{+}P_{2}^{+}}b^{\perp }\right] \,,
\end{equation}
where
\begin{equation}
\tilde{s}_{7}\left( b^{\perp }\right) =\int d^{2}P^{\perp }\exp \left[
-i(b^{\perp }P^{\perp })\right] s_{7}(P^{\perp })
\end{equation}
and
\[
\tilde{F}\left( x,\xi ,b^{\perp }\right) =\int \frac{d^{2}\tilde{\Delta}
^{\perp }}{(2\pi )^{2}}\exp \left[ i(\tilde{\Delta}^{\perp }b^{\perp })
\right] 
\]
\[
\times H\left( x,\xi ,-\frac{|\tilde{\Delta}^{\perp }|^{2}+4\xi ^{2}M^{2}}{
1-\xi ^{2}}\right) 
\]
\begin{equation}
=\int \frac{d^{2}\tilde{\Delta}^{\perp }}{(2\pi )^{2}}\exp \left[ i(\tilde{
\Delta}^{\perp }b^{\perp })\right] F(x,\xi ,\tilde{\Delta}^{\perp })\,.
\label{F-impact-def}
\end{equation}
Now we find from the inequality (\ref{ineq-1})
\[
\int d^{2}b^{\perp }|\tilde{s}_{7}\left( b^{\perp }\right) |^{2}\int
dP_{1}^{+}dk_{1}^{+}\int dP_{2}^{+}dk_{2}^{+}\frac{(P_{1}^{+}+P_{2}^{+})^{2}
}{P_{1}^{+}P_{2}^{+}} 
\]
\[
\times s_{6}(k_{1}^{+},P_{1}^{+})s_{6}^{\ast }(k_{2}^{+},P_{2}^{+})2\pi
\delta (P_{1}^{+}-k_{1}^{+}-P_{2}^{+}+k_{2}^{+}) 
\]
\begin{equation}
\times \tilde{F}\left[ \frac{k_{1}^{+}+k_{2}^{+}}{P_{1}^{+}+P_{2}^{+}},\frac{
P_{1}^{+}-P_{2}^{+}}{P_{1}^{+}+P_{2}^{+}},\frac{P_{1}^{+}+P_{2}^{+}}{
2P_{1}^{+}P_{2}^{+}}b^{\perp }\right] \geq 0\,.
\end{equation}
Here function $\tilde{s}_{7}\left( b^{\perp }\right) $ is arbitrary.
Therefore this inequality should hold before the integration over $b^{\perp
} $

\[
\int\limits_{0}^{\infty
}dP_{1}^{+}\int\limits_{0}^{P_{1}^{+}}dk_{1}^{+}\int\limits_{0}^{\infty
}dP_{2}^{+}\int\limits_{0}^{P_{2}^{+}}dk_{2}^{+}s_{6}(k_{1}^{+},P_{1}^{+}) 
\]
\[
\times s_{6}^{\ast }(k_{2}^{+},P_{2}^{+})\tilde{F}\left[ \frac{
k_{1}^{+}+k_{2}^{+}}{P_{1}^{+}+P_{2}^{+}},\frac{P_{1}^{+}-P_{2}^{+}}{
P_{1}^{+}+P_{2}^{+}},\frac{P_{1}^{+}+P_{2}^{+}}{2P_{1}^{+}P_{2}^{+}}b^{\perp
}\right] 
\]
\begin{equation}
\times \delta \left[ (P_{2}^{+}-k_{2}^{+})-(P_{1}^{+}-k_{1}^{+})\right] 
\frac{(P_{1}^{+}+P_{2}^{+})^{2}}{P_{1}^{+}P_{2}^{+}}\geq 0\,.  \label{ineq-2}
\end{equation}
The integration limits are taken from Eq.~(\ref{s-region}).

The inequality (\ref{ineq-2}) should hold for any value of $b^{\perp }$ and
for any function $s_{6}(k^{+},P^{+})$. The next step is to take 
\begin{equation}
s_{6}(k^{+},P^{+})=s_{8}(P^{+}-k^{+})s_{9}(P^{+})\,.
\end{equation}
In the limit 
\begin{equation}
\left| s_{8}(u)\right| ^{2}\rightarrow \delta (u-v)\,,\quad v>0
\end{equation}
we find from inequality (\ref{ineq-2})
\[
\int\limits_{0}^{\infty
}dP_{1}^{+}\int\limits_{0}^{P_{1}^{+}}dk_{1}^{+}\int\limits_{0}^{\infty
}dP_{2}^{+}\int\limits_{0}^{P_{2}^{+}}dk_{2}^{+}s_{9}(P_{1}^{+})s_{9}^{\ast
}(P_{2}^{+}) 
\]
\[
\times \tilde{F}\left[ \frac{k_{1}^{+}+k_{2}^{+}}{P_{1}^{+}+P_{2}^{+}},\frac{
P_{1}^{+}-P_{2}^{+}}{P_{1}^{+}+P_{2}^{+}},\frac{P_{1}^{+}+P_{2}^{+}}{
2P_{1}^{+}P_{2}^{+}}b^{\perp }\right] 
\]
\[
\times \delta \left[ (P_{1}^{+}-k_{1}^{+})-v\right] \delta \left[
(P_{2}^{+}-k_{2}^{+})-v\right] \frac{(P_{1}^{+}+P_{2}^{+})^{2}}{
P_{1}^{+}P_{2}^{+}} 
\]
\[
=\int\limits_{v}^{\infty }dP_{1}^{+}\int\limits_{v}^{\infty
}dP_{2}^{+}s_{9}(P_{1}^{+})s_{9}^{\ast }(P_{2}^{+})\frac{
(P_{1}^{+}+P_{2}^{+})^{2}}{P_{1}^{+}P_{2}^{+}} 
\]
\begin{equation}
\times \tilde{F}\left[ \frac{P_{1}^{+}+P_{2}^{+}-2v}{P_{1}^{+}+P_{2}^{+}},
\frac{P_{1}^{+}-P_{2}^{+}}{P_{1}^{+}+P_{2}^{+}},\frac{P_{1}^{+}+P_{2}^{+}}{
2P_{1}^{+}P_{2}^{+}}b^{\perp }\right] \geq 0\,.  \label{ineq-2b}
\end{equation}
We change from $P_{1}^{+},P_{2}^{+}$ to the new integration variables
\begin{equation}
x=\frac{P_{1}^{+}+P_{2}^{+}-2v}{P_{1}^{+}+P_{2}^{+}}\,,\quad \xi =\frac{
P_{1}^{+}-P_{2}^{+}}{P_{1}^{+}+P_{2}^{+}}
\end{equation}
\begin{equation}
P_{1}^{+}=v\frac{1+\xi }{1-x}\,,\quad P_{2}^{+}=v\frac{1-\xi }{1-x}
\end{equation}
with the Jacobian
\begin{equation}
\frac{D(P_{1}^{+},P_{2}^{+})}{D(\xi ,x)}=\frac{2v^{2}}{(1-x)^{3}}\,.
\end{equation}
After this change of variables we have
\begin{equation}
\frac{(P_{1}^{+}+P_{2}^{+})^{2}}{4P_{1}^{+}P_{2}^{+}}=\frac{1}{1-\xi ^{2}}\,.
\end{equation}
Then inequality (\ref{ineq-2b}) takes the form
\[
\int\limits_{-1}^{1}d\xi \int\limits_{|\xi |}^{1}dx\frac{1}{(1-x)^{3}(1-\xi
^{2})}s_{9}^{\ast }\left( v\frac{1-\xi }{1-x}\right) 
\]
\begin{equation}
\times s_{9}\left( v\frac{1+\xi }{1-x}\right) \tilde{F}\left( x,\xi ,\frac{
(1-x)b^{\perp }}{v(1-\xi ^{2})}\right) \geq 0\,.  \label{ineq-3}
\end{equation}
Since function $s_{9}$ and parameter $b^{\perp }$ are arbitrary, we can set $
v=1$ without losing generality. With $s_{9}(z)=zp(z^{-1})$ we obtain
\[
\int\limits_{-1}^{1}d\xi \int\limits_{|\xi |}^{1}\frac{dx}{(1-x)^{5}}p^{\ast
}\left( \frac{1-x}{1-\xi }\right) p\left( \frac{1-x}{1+\xi }\right) 
\]
\begin{equation}
\times \tilde{F}\left( x,\xi ,\frac{1-x}{1-\xi ^{2}}b^{\perp }\right) \geq
0\,.
\end{equation}
Thus the inequality (\ref{ineq-res}) with the upper $\pm $  sign is
established. The case of the other sign can be considered in a similar way.


\begin{thebibliography}{99}
\bibitem{MRGDH-94} D.~M\"{u}ller, D.~Robaschik, B.~Geyer, F.-M.~Dittes, and
J.~Ho\v{r}ej\v{s}i, Fortschr.\ Phys.\ \textbf{42} (1994) 101.

\bibitem{Radyushkin-96-a} A. V.~Radyushkin, Phys.\ Lett.\ \textbf{B380}
(1996) 417.

\bibitem{Radyushkin-96} A. V.~Radyushkin, Phys.\ Lett.\ \textbf{B385} (1996)
333.

\bibitem{Ji-97} X.~Ji, Phys.\ Rev.\ Lett.\ \textbf{78} (1997) 610.

\bibitem{Ji-97-b} X.~Ji, Phys.\ Rev.\ \textbf{D55} (1997) 7114.

\bibitem{CFS-97} J. C.~Collins, L.~Frankfurt, and M.~Strikman, Phys.\ Rev.\ 
\textbf{D56} (1997) 2982.

\bibitem{Radyushkin-97} A. V.~Radyushkin, Phys.\ Rev.\ \textbf{D56} (1997)
5524.

\bibitem{Radyushkin-review} A. V.~Radyushkin, in Shifman M. (ed.): At the
frontier of particle physics, vol. 2, pp. 1037-1099 (World Scientific,
Singapore, 2001). 

\bibitem{GPV} K.~Goeke, M. V.~Polyakov, and M.~Vanderhaeghen, Prog.\ Part.\
Nucl.\ Phys.\ \textbf{47} (2001) 401. 

\bibitem{BMK-2001} A.V.~Belitsky, D.~M\"{u}ller, and A.~Kirchner, Nucl.
Phys. \textbf{B629} (2002) 323.

\bibitem{AMS} A. Freund, M. McDermott, and M. Strikman, hep-ph/0208160.

\bibitem{Martin-98} A. D.~Martin and M. G.~Ryskin, Phys.\ Rev.\ \textbf{D57}
(1998) 6692.

\bibitem{Radyushkin-99} A. V.~Radyushkin, Phys.\ Rev.\ \textbf{D59} (1999)
014030.

\bibitem{PST-99} B.~Pire, J.~Soffer, and O.~Teryaev, Eur. Phys. J. \textbf{C8%
} (1999) 103.

\bibitem{Ji-98} X.~Ji, J. Phys. \textbf{G24} (1998) 1181.

\bibitem{DFJK-00} M.~Diehl, T.~Feldmann,~R. Jakob, and P.~Kroll, Nucl. Phys. 
\textbf{B596} (2001) 33.

\bibitem{Burkardt-01} M.~Burkardt, hep-ph/0105324.

\bibitem{Pobylitsa-01} P. V. Pobylitsa, Phys. Rev. \textbf{D65} (2002) 
077504.

\bibitem{Pobylitsa-02} P. V.~Pobylitsa, Phys. Rev. \textbf{D65} (2002)
114015. 

\bibitem{Diehl-02} M.~Diehl, Eur. Phys. J. \textbf{C25} (2002) 223. 

\bibitem{Burkardt-02-a} M.~Burkardt, Nucl. Phys. \textbf{A711} (2002) 127. 

\bibitem{Burkardt-02-b} M.~Burkardt, hep-ph/0207047.

\bibitem{Pobylitsa-02-c} P. V. Pobylitsa, Phys. Rev. \textbf{D66} (2002)
094002. 

\bibitem{Burkardt-00} M.~Burkardt, Phys. Rev. \textbf{D62} (2000) 071503.

\bibitem{Pobylitsa-02-e} P. V. Pobylitsa, hep-ph/0210238.

\bibitem{Pobylitsa-02-d} P. V. Pobylitsa, hep-ph/0210150.

\bibitem{TM-02} B. C.~Tiburzi and G. A.~Miller, hep-ph/0209178.

\bibitem{BR-00} J. Bl\"{u}mlein and D. Robaschik, Nucl. Phys. \textbf{B581}
(2000) 449. 
\end{thebibliography}
\end{document}